\documentclass[12pt,reqno]{amsart}

\usepackage{mathrsfs}
\usepackage{multicol}
\usepackage{epsfig}
\usepackage{latexsym}
\usepackage{amsmath}
\usepackage{enumerate}
\usepackage{amssymb}
\usepackage{amsfonts}
\usepackage{graphicx}
\usepackage{amscd}
\usepackage{verbatim}
\usepackage{calc}
\begin{document}

\begin{center}
\textbf{Linearization Stability of Einstein Field Equations is a Generic Property}
\end{center}
	
\begin{center}
R.V.Saraykar\\
Department of Mathematics, RTM Nagpur University, University Campus, Nagpur-440033, India\\
Mail : ravindra.saraykar$@$gmail.com\vspace{0.5cm}\\

Juhi H. Rai \\
Department of Mathematics, RTM Nagpur University, University Campus, Nagpur-440033, India \\
Mail : juhirai$28@$gmail.com  \\

\end{center}
 \textbf{Abstract}: In this paper, we prove that Linearization Stability of Einstein Field Equations is a Generic Property in the sense that within the class $\mathcal{V}$ of space-times which admit a compact Cauchy hypersurface of constant mean curvature, the subclass $\mathcal{V}_K$ of space-times which are linearization stable forms an open and dense subset of $\mathcal{V}$ under $C^\infty$ - topology. The work is based upon the work of Ebin $[1]$,  Fischer, Marsden and Moncrief $[2,3]$ and Beig, Chrusciel and Schoen $[4]$. 

\begin{enumerate}[1.]
\item \textbf{INTRODUCTION} \vspace{5pt}\\
The issue of Genericity of a property plays an equally important role as the issue of stability in Mathematics, Mathematical and Physical sciences, and other sciences also.\\ 
When we define a certain property of a topological space, or a measure space, or property of a dynamical system or that of a space-time in the theory of relativity, we expect that such a property will hold for "almost all" objects.  The word "generic" is used to describe such properties. Thus, properties that hold for "typical" examples are called generic properties. For instance, "a generic polynomial does not have a root at zero," or "A generic matrix is invertible." As another example, "If $f:{M}\rightarrow {N}$ is a smooth function between smooth manifolds, then a generic point of  $N$ is not a critical value of  $f$." (This is by Sard's theorem.)\\
There are many different notions of "generic" (what is meant by "almost all") in mathematics, with corresponding dual notions of "almost none" (negligible set). The two main classes are:\\
( I ) In measure theory, we say that a property is generic if it holds almost everywhere. This means that a property holds for all points except for a set of measure zero.\\
( II ) In topology and algebraic geometry, a generic property is one that holds on a dense open set, or more generally on a residual set, with the dual concept being a nowhere dense set, or more generally a meagre set.\\
In discrete mathematics, one uses the term "almost all"  to mean cofinite (all but finitely many), cocountable (all but countably many), for sufficiently large numbers, or, sometimes, asymptotically almost surely. The concept is particularly important in the study of random graphs.\\
As mentioned above, in topology and algebraic geometry, a generic property is one that holds on a dense open set, or more generally on a residual set (a countable intersection of dense open sets), with the dual concept being a closed nowhere dense set, or more generally a meagre set (a countable union of nowhere dense closed sets). Same definition is used in the theory of Dynamical Systems also.
However, "denseness" alone is not sufficient to characterize a generic property. This can be seen even in the real numbers, where both the rational numbers and their complement, the irrational numbers, are dense in the set of real numbers. Since it does not make sense to say that both a set and its complement exhibit typical behavior, both the rationals and irrationals cannot be examples of sets large enough to be typical. Consequently we rely on the stronger definition above which implies that the irrationals are typical and the rationals are not. We also note that none of these sets is an open subset of real numbers under usual topology.\\
For applications, if a property holds on a residual set, it may not hold for every point, but perturbing it slightly will generally land one inside the residual set, and these are thus the most important case to address in theorems and algorithms.
For function spaces, a property is generic in $C^r{(M, N)}$ if the set holding this property contains a residual subset in the $C^r$ topology. Here $C^r{(M, N)}$ is the function space whose members are continuous functions with $r$ continuous derivatives from a manifold M to a manifold N. This topology is often used in the mathematics literature and is well-known as Whitney-$C^r$ topology.
We also note that the space $C^r{(M, N)}$ of $C^r$ mappings between $M$ and $N$  is a Baire space and hence any residual set is dense. This property of the function space is what makes generic properties typical.\\
In Diferential geometry, Ebin $[1]$ in his Ph.D. thesis in 1967, and in his subsequent article in 1970, studied the space $\mathcal{M}$ of Riemannian metrics on a compact manifold of dimension $ > 1$ and endowed it with a topology and differential structure induced by that on the space of mappings between two manifolds. In this work, he proved that the subset of the space  $\mathcal{M}$ consisting of metrics with trivial isometry group forms an open and dense subset of  $\mathcal{M}$. In this sense, this property is generic. As we shall see later, this property is not sufficient to ensure our result on linearization stability.\\
In general theory of relativity, there are few properties of a space-time which are known to be generic. For example, Hawking $[5]$ remarks that a property of a space-time being stably causal is a generic property. Another generic property is the "generic condition" used in Hawking-Penrose singularity theorems ( see for example, Hawking and Ellis $[6]$). Beem and Harris $[7]$ proved this result in mathematical detils. In the present short paper, we show that "linearization stability" of Einstein Field Equations is a Generic Property. More specifically, we prove that in the class $\mathcal{V}$ of space-times which admit a compact Cauchy hypersurface of constant mean curvature, the subclass $\mathcal{V}_K$ of space-times which are linearization stable forms an open and dense subset of $\mathcal{V}$ under $C^\infty$ - topology.
In classical general relativity, Einstein field equations for vacuum space-times are described by \\
$Ein(^4{g})$ = $^4R_{\mu\nu}-\frac{1}{2} ^4R g_{\mu\nu}=0$ where \\
$^4g_{\mu \nu}$ is space-time metric and $Ein(^4{g})$, $^4R_{\mu \nu}$ and $^4R$ are respectively the Einstein tensor, Ricci tensor and curvature scalar with respect to $^4g_{\mu \nu}$. If we consider a spacetime resulting from evolution of a three dimensional spacelike hypersurface $M$ which is usually taken as three dimensional compact or non-compact Riemannian manifold, then spacetime can be described as $M\times R$ and above field equations can be split into four constraint equations and six evolution equations in terms of three dimensional quantities defined on $M$. This is well-known as ADM formalism (see, for example, Misner,Thorne and Wheeler $[8]$, Chapter $21$) and splitting uses Gauss-Codazzi equations from differential geometry. These equations are given as follows :\\
Constraint equations:\\
\begin{equation*}
\begin{split}
\Phi_0(g,\pi)=& R(g)\sqrt{g}-(|\pi|^2-1/2(tr_g \pi)^2)/\sqrt{g}= 0,\\
\end{split}
\end{equation*}
(This is known as Hamiltonian constraint equation), and
\begin{equation*}
\begin{split}
\Phi_i(g,\pi)=& 2(\nabla^jK_{ij}-\nabla_i (tr_gK))\sqrt{g}=0\\
\end{split}
\end{equation*}
(This is known as Momentum constraint equation). \\

Here, $g$ is the sufficiently smooth Riemannian metric  on $M$ induced by $^4g_{\mu \nu}$, $K$ denotes the second fundamental form corresponding to $g$,  $\pi'=(K-(tr\ K)g)$ and $\pi=\pi'\otimes \mu_g\,$ where $\mu_g$ is the volume element corresponding to $g$. $\pi$ is momentum density conjugate to $g$ and $(g,\pi)$ form vacuum initial data in ADM ( Hamiltonian) formalism from which space-time evolves.\\
 
Evolution equations are given as follows: \\
\begin{equation*}
\begin{split}
\partial g/\partial t=& 2N[\pi'-\frac{1}{2}(tr\pi')g]-L_Xg \\
\partial \pi/\partial t=& NS_g(\pi,\pi)-[N\ \text{Ein(g)-Hess}\ N-g\triangle N]\mu_g-L_X\pi \\
\end{split}
\end{equation*}

Here $N$ is the Lapse function and $X$ is the shift vector field, \\
$ \triangle N = -g^{ij}  N_{|i|j} ,  HessN =N_{|i|j} , 
\\ Ein(g)=Ric(g) -\frac{1}{2} R(g)g  ,\\ 
S_g (\pi,\pi)= -2 [\pi'\times\pi' - \frac{1}{2} (tr \pi')\pi'] \mu_g +\frac{1}{2} g^\sharp [\pi'.\pi' - \frac{1}{2} (tr\pi')^2] \mu_g, and \\ (\pi'\times\pi')^{ij} = (\pi')^{ik} (\pi')^j_k  ,  \pi'.\pi' = (\pi')^{ij} (\pi')_{ij} $.\\
 We consider the Constraint function $\Phi=(\Phi_0,\Phi_i)=\Phi(g,\pi).$ \\
Mathematical aspects of this formalism such as the problem of linearization stability and its relationship with the presence of Killing fields, manifold structure of set of solutions of constraint equations and existence and uniqueness of solutions of constraint equations for vacuum spacetimes as well as spacetime with matter fields such as electromagnetic fields, Yang-Mills fields, scalar fields etc. attracted attention of mathematicians and theoretical physicists for more than four decades. These aspects are aptly described in the review articles by Fischer and Marsden $[2]$, Choquet-Bruhat and York $[9]$ and papers by Fischer, Marsden and Moncrief $[3]$, Arms $[10,11]$ , Saraykar and Joshi $[12]$ and Saraykar $[13,14]$,  Isenberg $[15]$ and Choquet-Bruhat, Isenberg and Pollock $[16]$. In the present paper, we restrict our attention to vacuum space-times for simplicity. As far as coupled matter fields are concerned, we make some remarks at the end of the paper. \\
Thus, in Section $2$, we describe definition of 'linearization stability' of vaccun Einstein field equations and the results on this aspect proved by Fischer, Marsden and Moncrief which are needed for our purpose. We also state our main theorem. In Section $3$, we give the proof of the main theorem and make some concluding remarks about Einstein field equations coupled with different matter fields and also about asymptotically flat space-times.\\
\item \textbf{Definition and results on Linearization Stability} \\
In this section we give the basic definitions and results on linearization stability. This section is based on the work by Fischer, Marsden and Moncrief [2,3]\\
We begin with the following definition :\\
Definition : Let $\Phi: X \longrightarrow Y$ be a non-linear differential operator between Banach spaces or Banach manifolds of maps $X$ and $Y$ (over a compact manifold ) . Consider the equation $\Phi(x)$ = $y_{0}$ for $y_{0} \in  Y$.\\
Let $ T_x{X} $ denote the tangent space to $ X $ at $x \in X$, and let \\ $D\Phi(x)$ : $T_x{X} \longrightarrow T_y{Y}$, with $y$ = $\Phi(x)$, be the Frechet derivative of $\Phi(x)$ at $x$. Thus to each solution $x_0$  of $D\Phi(x)$ = $y_0$ ,  $D\Phi(x_0).h = 0$, $h \in T_{x_{0}}X$, is the associated system of linearized equations about $x_0$, and a solution $h \in T_{x_0}X$ of linearized equations is an infinitesimal deformation ( or first order deformation ) of the solution $x_0$.
If for each solution $h$ of linearized equations , there exists a curve $x_t$   of exact solutions of $D\Phi(x) = y_0$  which is tangent to $h$ at $x_0$  i.e. $x(0) = x_0$  and $[(d / dt)(x_t )]_{|_{t = 0}} = h$, then we say that equation $\Phi(x) = y_{0}$ is linearization stable at $x_0$, and deformation $h$ is called integrable.\\
Useful criterion to prove linearization stability is as follows :\\
Theorem: Let $X$ and $Y$ be Banach manifolds, and $\Phi: X \longrightarrow Y$ be a $C^1$-map. Let $x_0 \in X$ be a solution of $\Phi(x) = y_{0}$. Suppose $D\Phi(x_0)$ is surjective with splitting kernel . Then the equation $\Phi(x)$ = $y_{0}$ is linearization stable at $x_0$.\\
( Splitting kernel means:  $T_{x}X = Range (D\Phi)^*(x) + Ker D\Phi(x)$). \\
Proof uses the Implicit Function Theorem.\\
We now discuss Linearization stability of Einstein's equations in case where Cauchy (spacelike) hypersurface is a compact 3 - manifold without boundary. 
In analogy with above definition, we define linearization stability of Einstein field equations :\\
We write Einstein equations for vacuum space-time as : $Ein(^4{g}) = 0$, where Ein denotes Einstein tensor. Let $^4{g}_0$ be a solution of Einstein equations. Then linearized equation is given by $DEin(^4{g}_0).^4{h}  = 0$. If for every $^4{h}$ satisfying linearized Einstein equation, there is a curve ($^4{g}_ {t}$) of exact solutions such that $[^4{g}_t]_{|t = 0}  = ^4{g_0}$ , and $[(d / dt)(^4{g_t} )]_{|_{t = 0}} =  ^4h$, then $^4{g}_0$ is called linearized stable.  In this case $^4{ h}$  is called integrable.
Not every $^4{ h}$ satisfying Linearized equation is integrable:
Brill and Deser [17] considered the space-time $({T^3})\times R$ and showed that there are solutions of linearized equation which are not integrable.
For more details, we refer the reader to the review article by Fischer and Marsden $[2]$.\\
In proving our main theorem, we shall need the following important results about Linearization Stability which are proved in $[2]$. We state them one by one:\\
Theorem $1$ : The Einstein system , defined by the evolution equations and constraint equations described above can be written as \\
(evolution equation) $\frac{\partial}{\partial\lambda} \begin{pmatrix} g \\
\pi \end{pmatrix} = J\cdot (D\Phi(g,\pi))^* \cdot\begin{pmatrix} N \\ X \end{pmatrix} $ \\where $J$ is the complex structure and $N$ and $X$ are the Lapse function and the shift vector field.\\
Let $\mathcal{C}_H$ = $\{(g,\pi): \Phi_0(g,\pi)= 0\}$ denote the set of solutions of Hamiltonian constraint, \\
and $\mathcal{C}_\delta$ = $\{(g,\pi): \Phi_i(g,\pi)= 0\}$ denote the set of solutions of momentum constraint. Thus, $\mathcal{C}$ = $\mathcal{C}_H \bigcap\mathcal{C}_\delta$ is the constraint set for Einstein system. \\
Moreover, we assume the following conditions :\\
$C_H $: If $\pi = 0 $, then $g$ is not flat;\\
$C_\delta$ :If  for $ X\in \chi(M), L_Xg = 0$ and $L_X \pi = 0$ , then $X=0$  \\
$C_{tr}$ : tr$\pi^\prime$ is a constant on M.\\
Then we have the following theorem : \\

\textbf{Theorem $2$:} Let $(g, \pi)\in\mathcal{C}_H \bigcap\mathcal{C}_\delta$ satisfy the conditions $C_H $, $C_\delta$ and $C_{tr}$ given above. Then the constraint set $\mathcal{C}$ = $\mathcal{C}_H \bigcap\mathcal{C}_\delta$ is a $C^\infty$ submanifold of $T^*M$ in a neighborhood of (g, $\pi$)with tangent space \\ $T_{(g,\pi)}\mathcal {C} = ker D\Phi(g,\pi)$.\\
Then, we have the following basic theorem on linearization stability.\\
\textbf{Theorem $3$:} Let $\Phi :T^*M\rightarrow C^\infty_d \times\Lambda^1_d $ and so $\mathcal{C} = \Phi^{-1}(0)$.\\ Let $(g_0 , \pi_0)\in\mathcal{C}$. Then the following conditions are equivalent: \\i)The constraint equations $\Phi(g,\pi)=0$ are linearization stable at $(g_0,\pi_0)$,\\ii)$D\Phi(g_0,\pi_0):S_2 \times S_d^2 \rightarrow C^\infty_d \times \Lambda^1_d $ is surjective,
\\iii)$D\Phi{(g_0,\pi_0)}^*:C^\infty \times \chi \rightarrow S^2_d X S_2$ is injective.\\
Here, $S_2$ denotes the class of symmetric $2$- tensors on $M$, $S^2_d$ denotes symetric $2$-tensor densities, $C^\infty$ denotes class of $C^\infty$- functions on $M$, $C^\infty_d$ denotes function densities, $\chi$ denotes $C^\infty$- vector fields on $M$ and $\Lambda^1$ and $\Lambda^1_d$ denote respectively one-forms and one-form densities on $M$. The spaces $C^\infty_d \times \Lambda^1_d$ and $C^\infty \times \chi$ are dual spaces of each other with natural $L^2$- pairing giving the appropriate inner product structure. Same is true for symmetric $2$-tensors and tensor densities.\\
The following theorems describe more results on linearization stability, and the isomorphism relationship between the set of independent Killing fields that a space-time admits, and the kernel of $D\Phi{(g_0,\pi_0)}^*$. It is noteworthy that the conditions which facilitate the property of linearization stability to hold, also makes the operator $D\Phi{(g_0,\pi_0)}^*$ elliptic. This makes further analysis simple because kernel of an elliptic operator is always finite dimensional. Thus we have : \\    
\textbf{Theorem $4$:} Let $(V, ^{(4)}g_0)$ be a vacuum spacetime which is the maximal development of Cauchy data $(g_0,\pi_0)$ on a compact hypersurface $M$.\\Then the Einstein equations  Ein $(^{(4)}g)=0 $ on $V$ are linearization stable at 
$(^{(4)}g_0)$ if and only if constraint equations $\Phi(g,\pi)=0$ are linearization stable at $(g_0,\pi_0)$ .\\In particular , if conditions $C_H $, $C_\delta$ and $C_{tr} $ hold for $(g_0,\pi_0)$ , then the Einstein equations are linearization stable.\\
\textbf{Theorem $5$:} Let $(^{(4)}g)$ be a solution to the empty-space field equations Ein $(^{(4)}g)=0 $. Let $M$ be a compact Cauchy hypersurface with induced metric $g_0$ and canonical momentum $\pi_0$ . Then ker $ D\Phi{(g_0,\pi_0)}^*$ (a finite-dimensional vector space) is isomorphic to the space of Killing vector fields of $(^{(4)}g)$ . In fact, $(N,X)\in ker D\Phi{(g_0,\pi_0)}^*$ if and only if there exists a Killing vector field $^4X$ of $(^{(4)}g)$ whose normal and tangential components to $M$ are $N$ and $X$.\\
\textbf{Theorem $6$:} Let $^{(4)}g_0$ be a solution of the vacuum field equations Ein $(^{(4)}g)=0 $ on $V$. Assume that $(V, ^{(4)}g_0)$ has a compact Cauchy surface $M$ and that  $(V, ^{(4)}g_0)$ is the maximal development. Then the Einstein equations on 
$V$, Ein $(^{(4)}g)=0 $ , are linearization stable at $^{(4)}g_0$ if and only if $^{(4)}g_0$ has no killing vector fields.\\

Thus, summarizing, for a space-time $(V, ^4{g})$ admitting a compact constant mean curvature (CMC) Cauchy hypersurface $M$, $(D\Phi)^*(g,\pi)$ is elliptic, 
the space of Killing fields of $^4{g}$  is isomorphic to the kernel of $(D\Phi)^*(g,\pi)$, and if $(V, {^4{g}})$  has no Killing fields, then it is linearization stable and conversely, if $(V, ^4{g})$ is linearization stable, then ker $(D\Phi)^*(g,\pi)$ is trivial.\\
Moreover, since kernel of an elliptic operator is finite dimensional, a space-time admitting compact CMC hypersurfaces can admit only a finite number of independent Killing fields. It will not be out of place to mention that similar results hold for Einstein field equations coupled with matter fields such as scalar fields, electro-magnetic fields and Yang-Mills fields. See, for example, $[10-12]$.
For full analysis of the structure of the space of solutions of Einstein's field equations in the presence of Killing fields, we refer the reader to Fischer, Marsden and Moncrief $[3]$ and Arms, Marsden and Moncrief $[18]$. See also Saraykar $[14]$ for scalar fields case.\\
We now state our main theorem : \\
\textbf{Theorem} : Let $\mathcal{V}$ denote the class of space-times which admit a compact Cauchy hypersurface of constant mean curvature. Then the subclass $\mathcal{V}_K$ of space-times which are linearization stable form an open and dense subset of $\mathcal{V}$ under $C^\infty$ - topology ( in general under Whitney-$C^r$ - topology). Thus, in this sense, linearization stability is a generic property.\\

\item \textbf{Proof of the main theorem} \\
We now consider the class $\mathcal V$ of all vacuum space-times ( equivalently class of Lorentz metrics satisfying Einstein field equations for vacuum space-times ) possessing compact Cauchy hypersurfaces of constant mean curvature (CMC). Such a class of space-times can be endowed with a suitable topology whose choice can be made as per our need. Such topologies have been discussed by Hawking $[5]$,Lerner $[19]$ and Beig, Chrusciel and Schoen $[4]$. For this class of space-times, we have theorems $1-6$ as proved in Fischer and Marsden $[2]$, which we have stated in the previous section. We use them in the following argument:\\
Thus, as noted above,space-times admitting compact CMC hypersurfaces can admit only finite number of independent Killing fields. Also,such a space-time $(V, {^4{g}})$ is linearization stable if and only if it has no Killing fields if and only if ker $(D\Phi)^*(g,\pi)$ is trivial.Any vector field $^4X$ on a space-time can be identified with a pair $(N,X)$ where $N$ is the perpendicular component of $^4X$ and $X$ is its parallel (space-like)component. This decomposition is of course related to Riemannin metric $g$ on the Cauchy hypersurface $M$. Beig et.al call this pair $(N,X)$ as Killing Initial Data (KID) when $^4X$ is a Killing field on $(V, {^4{g}})$, and $(g,\pi)$ is called vacuum initial data, where $(g,\pi)$ is as in ADM formalism explained above. Thus, each space-time $(V, {^4{g}})$ corresponds to a vacuum initial data $(g,\pi)$ and the Killing field $^4X$ which a space-time admits, corresponds to a KID $(N,X)$. Due to covariance property of Einstein field equations, Killing property of initial data will be carried throughout the evolution. Thus this property is possessed by space-time as a whole. In other words, KIDs are in one-to-one correspondence with Killing vectors in the associated space-time. With this association,and keeping above result in mind, it is now clear that the class of vacuum space-times $\mathcal{V}_K$ possessing a compact cauchy hypersurface which are linearization stable can be identified with the class $\mathcal{V}_I$  of all vacuum initial data $(g,\pi)$ without (global) KID with $tr(\pi)$ = constant. To this class $\mathcal{V}_I$, we now apply the following theorem proved by Beig, Chrusciel and Schoen $[4]$ :\\
\textbf{Theorem:}(Theorem $(1.3)$ of $[4]$ ): Consider the following collections of vacuum initial data sets:\\
1.$\Lambda=0$ with an asymptotically flat region ,or\\
2. $tr_gK=\Lambda=0$ with an asymptotically flat region ,or\\
3.$K=\Lambda=0$with an asymptotically flat region ,or\\
4.with a conformally compactifiable region in which$tr_gK$ is constant , or\\
5. the trace of $K$ is constant and the underlying manifold M is compact , with\\
${(tr_gK)}^2\geq\frac{2n}{n-1} \Lambda$, or\\
6.$K=0$ , M is compact, and the curvature scalar R satisfies $R=2\Lambda\leq0$, with a $ C^{k,\alpha}\times C^{k,\alpha}$ weighted in the non-compact region) topology, with $k\geq k_0(n)$ for some $K_0(n) (k\geq$6  if  n=dimM=3). For each such collection the subset of vacuum initial data sets without global KID is open and dense.\\
Here $\Lambda$ is a cosmological constant. Since we are considering vacuum space-times, $\Lambda=0$, and hence if ${(tr_gK)}$ is constant, second condition in $(5)$ above is automatically satisfied. Of course, ${(tr_gK)}$ is constant is equivalent to ${(tr_g\pi)}$ is constant.\\
Beig, Chrusciel and Schoen $[4]$ also prove that above result remains true if Cauchy surface is asymptotically flat or asymptotically hyperbolic.\\
Thus, applying this theorem under condition $5$, we conclude that the class $\mathcal{V}_I$ is open and dense in $\mathcal{V}$. Openness and denseness is to be considered under appropriate topology, as explained in $[4]$. Thus, as explained above, we can now conclude that the class of space-times possessing compact Cauchy hypersurface of constant mean curvature which are linearization stable forms an open and dense subset of $\mathcal V$. In this sense, property of linearization stability is generic. This completes the proof of the theorem.\\
Some remarks are in order.\\ 

Remarks : \\
$1$. In support of this result, we quote a somewhat similar result proved by Mounoud $[20]$ :\\
Theorem (Theorem $1$, Mounoud $[20]$) : Let $V$ be a compact manifold with dimension $\geq 2$ and $M_{p,q}$ be the set of smooth pseudo-Riemannian metrics of signature $(p,q)$ on $V$. Then the set $G =\left\{ g \in  M_{p,q}: Is(g) = Id \right\} $ contains a subset that is open and dense in $M_{p,q}$  for the $C^\infty$-topology. Here, $Is(g)$ denotes isometry group of $g$. \\
$2$. As remarked in the Introduction, the first result of this kind for Riemannian metrics was proved by Ebin $[1]$ in 1970. He proved that the set of Riemannian metrics without isometries on a compact manifold is open and dense in the set of all Riemannian metrics under suitable topology. Similar but more general results were noted and proved by Henrique de A. Gomes $[21]$ by using Ebin-Palais slice theorem. However, these results do not imply directly the generic property of linearization stability of space-times under consideration, even though the condition that isometry group of $g$ is trivial is equivalent to saying that $g$ admits no non-trivial Killing field. This is because, vanishing of Lie derivative of $^4{g}$ with respect to $^4X$ involves vanishing of Lie derivatives of $g$ and $\pi$ both, and triviality of isometry group of $g$ alone does not imply that of $\pi$. This has been explained in more mathematical details in $[4]$. Thus, different argument is needed in proving Theorem $1.3$ in $[4]$, which enabled us to prove our result.\\  
$3$. Finally, in the context of general relativity, it is interesting to note that it is not yet fully known if the above generic result is valid when constant mean curvature condition is removed.\\

\item \textbf{Concluding Remarks :}\\
In this paper, we have proved that within the class $\mathcal{V}$ of space-times which admit a compact Cauchy hypersurface of constant mean curvature, the subclass $\mathcal{V}_K$ of space-times which are linearization stable forms an open and dense subset of $\mathcal{V}$ under $C^\infty$ - topology. Beig, Chrusciel and Schoen $[4]$ work with $C^k$ - topology, but we can as well work with $C^\infty$ - topology. Thus, in this sense, linearization stability is a generic property. For simplicity, we have considered vacuum space-times with compact Cauchy hypresurface of constant mean curvature. It is well-known that (cf. $[10-12]$) Einstein field equations coupled with matter fields for space-times possessing compact Cauchy surfaces can be written in the Hamiltonian formalism (ADM-formalism) and similar linearization stability analysis is valid for such field equations. Hence, genericity result will also hold for such coupled systems. Moreover, since linearization stability results hold for asymptotically flat space-times also (cf. Choquet-Bruhat $[22]$ and Saraykar $[13]$) and the theorem proved by  Beig, Chrusciel and Schoen $[4]$ is valid for such space-times, linearization stability will remain a generic property for such space-times too.\\
\end{enumerate}
\textbf{References}
\begin{enumerate}[{[1]}]
\item D.G. Ebin, The manifold of Riemannian metrics, Proc. Symp. Pure Maths. Vol. XVI,11-40, (1970), "On the space of Riemannian metrics", Ph.D. Thesis at MIT, 1967.\\
\item A. Fischer and J.E. Marsden , Topics in the dynamics of general relativity, in "Isolated gravitating systems in general relativity", Ed. J. Ehlers, Italian Physical Society (1979), 322-395
\item A.Fischer, J.E.Marsden  and V. Moncrief , The structure of the space of solutions of Einstein's equations.I. One Kiling field,Annales de la Institut H. Poincare, Section A, Vol.33 (2) (1980), 147-194
\item R. Beig, P.T. Chrusciel and Richard Schoen, KIDs are non-generic, Ann. Inst. H. Poincare, Vol. 6, 155-194 (2005) \\
\item S.W. Hawking, Gen. Rel. Grav. 1, 393 (1970).\\
\item S.W. Hawking  and G.F.R Ellis ., The large scale structure of space-time , Cambridge university press , Cambridge , 1973.\\
\item J.K. Beem and S.G. Harris, Gen. Rel. and Grav. 25(9),939-961(1993).\\
\item C. Misner, K. Thorne and J.A. Wheeler, Gravitation, Chapter 21, Freeman Press, Sanfransisco, 1972.\\
\item Y.Choquet-Bruhat  and J.W. York , The Cauchy Problem. In: Held A. (Ed.) General Relativity and Gravitation, I. Plenum Press, New York,(1980), 99-172.\\
\item J. Arms, Jour. Math. Phys. 18, 830-833(1977).\\
\item J. Arms, Jour. Math. Phys. 20, 443-453(1979).\\
\item R.V.Saraykar  and N.E.Joshi , Linearisation stability of Einstein's equations coupled with self-gravitating scalar fields, Jour. Math. Phys. Vol.22 no.2, (1981), 343-347; Erratum Vol.23, (1982), 1738.\\
\item R.V.Saraykar , Linearisation stability of coupled gravitational and scalar fields in Asymptotically flat space-times, Pramana, Vol.19, No.1,  (1982), 31-41.\\
\item R.V.Saraykar , The structure of the space of solutions of Einstein's equations coupled with Scalar fields, Pramana. Vol.20, No.4, (1983), 293-303.\\
\item J.Isenberg , "The initial value problem in general relativity", chapter in "The Springer Handbook of Spacetime," edited by A. Ashtekar and V. Petkov. (Springer-Verlag), (2014), 303-321 (arXiv:1304.1960v1 [gr-qc]).\\
\item Y.Choquet-Bruhat ,J. Isenberg  and D.Pollack , The constraint equations for the Einstein-scalar field system on compact manifolds. Classical Quantum Gravity, 24(4),(2007), 809–828.\\
\item D. Brill and S. Deser,  Commu. Math. Phys. 32, 291-304 (1973) \\
\item J. Arms ,J.E. Marsden   and V.Moncrief , The structure of the space of solutions of Einstein's equationsII. Several Kiling fields and the Einstein-Yang-Mills Equations, Annals of Physics, Vol. 144 (1) (1982), 81-106.\\
\item D.Lerner, Topology on the space of Lorentz metrics, Commu. Math. Phys., Vol.32 , 19 - 38, (1973).\\
\item P. Mounoud, Metrics without isometries are generic,  arXiv:1403.0182v1 [math.DG] \\
\item H. d. A. Gomes, A Note on the Topology of a Generic Subspace of Riem,  arXiv:0909.2208v1 [math-ph]\\
\item Y. Choquet-Bruhat , "General Relativity and The Einstein Field Equations", Oxford monographs in Physics, Clarendon Press, Oxford, 2009

\end{enumerate}

\end{document}